\documentclass{ws-ijmpd}
\usepackage[super,compress]{cite}
\usepackage{amsmath}
\usepackage{graphicx}
\usepackage{epstopdf}
\usepackage{color}
\usepackage{multicol}
\usepackage{aas_macros}
\usepackage{hyperref}

\newcommand{\PB}{\textsc{Polarbear}}
\newcommand{\bicep}{B{\sc icep}}

\begin{document}

\markboth{Kaufman, Leon, Keating}
{Using the Crab Nebula as a high precision calibrator for CMB polarimeters}

%
\catchline{}{}{}{}{}
%






\title{USING THE CRAB NEBULA AS A HIGH PRECISION CALIBRATOR FOR COSMIC MICROWAVE BACKGROUND POLARIMETERS}

\author{Jonathan Kaufman\footnote{jkaufman@physics.ucsd.edu}, David Leon, Brian Keating}
\address{Department of Physics, University of California, San Diego, 9500 Gilman Drive, La Jolla, CA 92093-0424, USA}

\maketitle

\begin{history}
\received{Day Month Year}
\revised{Day Month Year}
\end{history}

\begin{abstract}

The polarization of the Cosmic Microwave Background (CMB) provides a plethora of information about the early universe.  Most notably, gravitational waves from the Inflationary epoch (the leading explanation of the origin of the universe) create a unique CMB polarization $B$-mode signal.  An unambiguous detection of the inflationary $B$-mode signal would be a window into the physics of the universe as it was $10^{-36}$ seconds after the Big Bang, at energy scales many orders of magnitude larger than what the LHC can produce.  However, there are several instrumental and astrophysical sources that can obfuscate the inflationary $B$-mode signal.  One of the most difficult parameters to calibrate for CMB telescopes is the absolute orientation of the antenna's polarization sensitive axis.  A miscalibration of the polarization orientation rotates the much brighter $E$-mode signal, producing a false $B$-mode signal.  The current best uncertainty on polarization orientation in the CMB community is $0.5^\circ$, set from extrapolating IRAM measurements of the Crab Nebula supernova remnant at 90 GHz to 150 GHz, where the CMB signals peak.  This accuracy is not sufficient to convincingly detect $B$-modes predicted by currently allowable models of Inflation.  We suggest to precisely measure the Crab Nebula's polarization, which can be calibrated absolutely to $0.1^\circ$ from measurements of the polarized emission of Mars, and use these data to calibrate current and upcoming CMB experiments.  In addition to inflationary $B$-modes, more precise calibration will allow us to better constrain the sum of the neutrino masses and set limits on exotic physics such as parity violation through cosmic polarization rotation.
\end{abstract}


\ccode{PACS numbers:}

\section{Introduction}

The Cosmic Microwave Background (CMB) is the glowing remnant of the Big Bang that permeates the cosmos.  Formed when the universe was 380,000 years old and redshifted today to microwave wavelengths, it is a snapshot of the transition between the hot, dense, primordial plasma to the universe we currently observe.

The CMB is an extremely uniform 2.725 K blackbody with fluctuations on the order of a few parts per 10,000 at degree-scales.  These anisotropies result from slight over-densities and under-densities in the underlying dark matter concentrations that seeded the formation of large scale structure.  The leading explanation of the origin of these anisotropies comes from Inflation, the rapid expansion of space-time in the very early universe.  The Inflation epoch amplified quantum fluctuations in the inflaton field to macroscopic scales, seeding anisotropies in the energy density and sending gravitational waves rippling through space-time.\cite{HuWhite}  Though there is much evidence for a period of Inflation in our early universe, an unambiguous detection of the gravitational waves uniquely generated from this expansion would be irrefutable evidence.\cite{Lidsey1997}

In addition to its temperature anisotropy, the CMB is partially polarized.  When CMB photons Thomson scatter off of electrons in the primordial plasma, they will become polarized if the incident photons have a quadrupolar anisotropic distribution.  Different physical mechanisms that produce the quadrupolar anisotropy in the photon bath will produce different polarization patterns.  These polarization patterns are known as $E$-mode and $B$-mode based on their parity properties, with $E$-modes being parity even and $B$-modes being parity odd.  Well-understood plasma physics produces $E$-mode type polarization in the CMB, but only physics beyond the standard model can produce the parity-odd $B$-modes.\cite{HuWhite}  The gravitational waves created during Inflation will produce $B$-mode polarization.  Thus a detection of $B$-mode polarization is essentially a detection of Inflation, as well as an indirect detection of gravitational waves, a feat that was rewarded with the 1993 Nobel Prize in physics.\cite{HulseTaylor}  The level of Inflationary $B$-modes is described by the tensor-to-scalar ratio, $r$, which is related to the energy scale of Inflation.  Smaller $r$ corresponds to fainter $B$-mode polarization, and thus a lower energy scale of Inflation.

There are late-time foregrounds which can obfuscate the $B$-mode polarization.  For example massive galaxy clusters gravitationally lens the CMB photons, mixing the brighter $E$-modes into  $B$-modes\footnote{It will also mix $B$-modes into $E$-modes but this is a much smaller effect since the $B$-modes are significantly fainter.}.  This effect is separable from the Inflationary $B$-mode signal and is also rich in information about the makeup of the universe; in particular, it is very sensitive to the sum of the neutrino masses\cite{Abazajian2013} and is one of the primary goals of current generation and near-future CMB experiments like \PB\ and the Simons Array\footnote{In addition, dust grains caught in our Galaxy's magnetic fields emit polarized radiation that will obscure underlying $B$-mode polarization.  This effect can be separated with high precision measurements of the CMB at several frequencies.} \cite{PB2014}.

The CMB has been studied since the 1960s, however it wasn't until the early 2000s that the $E$-mode polarization of the CMB was measured \cite{DASI}.  Since then, increasingly more sensitive telescopes have been built and deployed on Earth, balloons, and in space, providing incredibly detailed $E$-mode polarization maps of the CMB.  With the March, 2014 detections of the faint $B$-mode polarization \cite{PB2014, B22014}, the race began to tease out the Inflationary gravitational wave signature. 

\section{Polarization Orientation Calibration}

In addition to astrophysical sources that can obscure the $B$-mode polarization, several instrumental errors can destroy or mimic a polarized signal in a systematic way.  One of the effects most difficult to calibrate is the absolute orientation of the polarization-sensitive axes of the telescope's antennas.  A small miscalibration of the polarization orientation will leak the brighter $E$-mode signal into the fainter $B$-mode signal as
\begin{equation}
C_{\ell,\text{leaked}}^{BB} \approx 4\alpha^2 C_\ell^{EE},
\label{eq:leaked_BB}
\end{equation}
where $\alpha$ is the difference between the true orientation angle and the calibrated orientation angle, and $C_\ell^{EE}$ is the CMB $E$-mode power spectrum.  The multipole, $\ell$, represents the angular scale, with a small $\ell$ corresponding to large separations between points on the sky and a large $\ell$ represents small separations on the sky.

Several methods have been used to calibrate the orientation angles, either in the lab or in the field, with varying levels of success \cite{Kaufman2013}.  However, there is no single, universal, terrestrial calibration method, and each method comes with a host of challenges. For example, for a telescope like \PB, the Fraunhofer distance is $\sim$ 4 km.  Even with mountainous local terrain, there is no suitable place at this distance to install a polarized microwave source that would be above the lower telescope's observable elevation limit.  Measurements at a closer distance are difficult due to the high loading on the bolometers (due to the low elevation required to observe them) as well as reflections off of the ground.  Furthermore, extrapolation of polarization orientation angles measured in the near-field to the far-field is daunting, and requires highly accurate physical optics modeling.  A stable, polarized astrophysical source would be an ideal calibrator.  Though there are many well studied polarized sources, we require a source that is both bright enough and large enough to be observed by the relatively course resolution of CMB telescopes (for example, the \PB\ beam size is 3.5$'$).

The Crab Nebula is an extended supernova remnant at a distance of $\sim$ 2 kpc  that emits highly polarized radiation in microwave frequencies \cite{Hester2008}.  Using the 30 meter IRAM telescope, the spatial polarization distribution at 90 GHz was measured with an absolute polarization accuracy of $0.5^\circ$ \cite{Aumont2009}.  This systematic error is five times larger than what can be achieved by the Very Large Array (VLA) using Mars as a polarization calibrator \cite{VLAMars}.  Although the flux is assumed to change with frequency as a power-law, the polarization angle has been shown by WMAP to be constant at the degree-level \cite{Weiland2011}.  High-precision polarization measurements of the Crab Nebula at radio and microwave frequencies would be an invaluable method for calibrating polarization angles of CMB telescopes.

The current achieved $0.5^\circ$ level of precision corresponds to a spurious $B$-mode signal as strong as $r = 0.01$ at 1$\sigma$ significance.  For current and future telescopes  that expect to probe this region of $r$, this level of uncertainty is unacceptable.  We note that the recent \emph{Planck} LFI and HFI results \cite{PlanckTauA2015} reinforce that the IRAM measurement of the Crab Nebula is the limiting factor for determining polarization orientation calibration for current CMB telescopes.

Using the radially symmetric polarization emission of Mars as a calibrator, a large radio telescope like the VLA can determine any frequency dependence of the Crab Nebula's polarization angle, and achieve a factor of five improvement on current polarization angle measurements \cite{VLAMars}.  Any CMB observatory with a view of the Crab Nebula, for example ACTPol, CLASS, \emph{Planck}, \PB, and the Simons Array, can then measure the spatial polarization angle distribution and compare it to these measurements as a high-precision calibrator.  

\begin{figure}[h]
\centering
\includegraphics[width=0.45\textwidth]{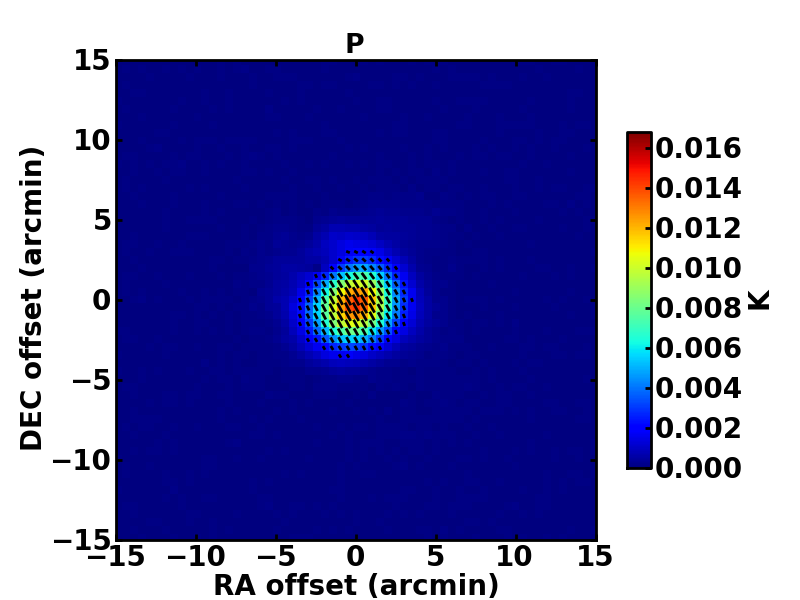}
\includegraphics[width=0.45\textwidth]{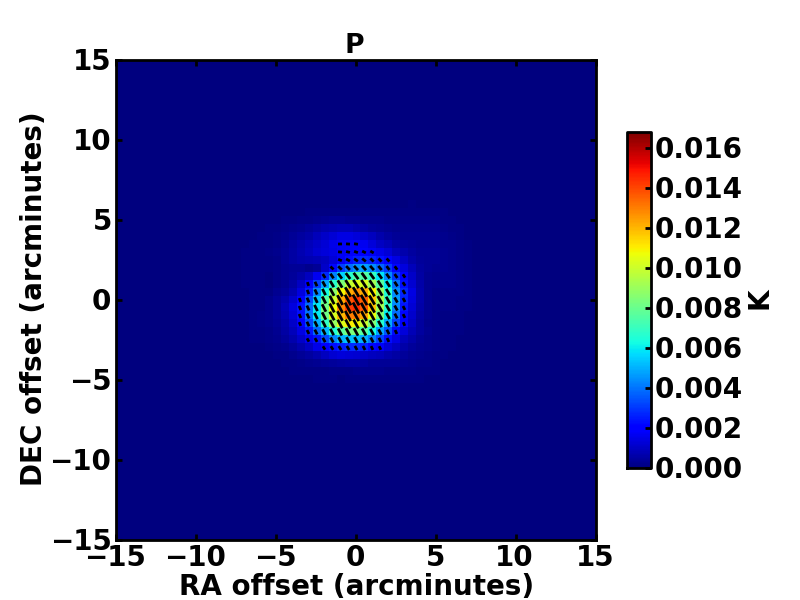}

\caption{Measured polarization map of the Crab Nebula at 150 GHz from the \PB\ experiment \cite{PB2014} (\emph{left}) and simulated \PB\ polarization map constructed from the 90 GHz IRAM measurement \cite{Aumont2009} (\emph{right}).  The IRAM data have a systematic uncertainty on the polarization angle of 0.5$^\circ$, five times worse than what can be achieved from measurements of Mars \cite{VLAMars}.} 
\label{fig:tauA}
\end{figure}

\section{Non-Inflationary Science Goals}

With precisely calibrated polarization orientation, a significant amount of exciting non-Inflationary fundamental physics can be probed with current and future CMB telescopes.

As mentioned above, the intervening large scale structure gravitationally lenses the CMB photons as they propagate through the universe.  This mixes $E$-modes into $B$-modes.  Since neutrinos wash out structure on small scales, these lensed $B$-modes are a very sensitive probe of the sum of the neutrino masses.  As with Inflationary $B$-modes, we require stringent calibration of the polarization orientation to avoid inducing spurious instrumental polarization.  The current systematic limit from IRAM measurements \cite{Aumont2009} is nearly at the level produced by lensing $B$-modes.

Due to the parity-odd nature of $B$-mode polarization, we can construct extremely sensitive probes of exotic physics by studying parity ``forbidden" correlations.  Correlations of the $B$-mode polarization with the scalar temperature anisotropies, $C_\ell^{TB}$, or with the parity-even $E$-mode polarization, $C_\ell^{EB}$, are expected to be zero\footnote{Up to the cosmic variance limit.}.  Cosmic polarization rotation, such as cosmic birefringence, can generate these correlations \cite{Kaufman2013, LueWangKamion1999, Feng2005}.

With parity violation having been demonstrated in the weak interaction \cite{Wu1957}, CP violation in the strong force, and the unification of the weak force with electromagnetism, we might expect that parity violation in the electromagnetic sector could exist at some level.  If we add the parity-violating and Lorentz invariance violating Chern--Simons term, $\mathcal{L}_{CS} = - \frac{1}{2} p_\mu A_\nu \tilde{F}^{\mu\nu}$, into the electromagnetic Lagrangian, and then solve the wave equation for this modified Lagrangian, different propagation speeds for left-handed and right-handed circularly polarized light in vacuum will result \cite{Carroll1990}.  For a linearly polarized photon traveling through vacuum, this would rotate the polarization plane by an angle, $\alpha$, dependent on the coupling four-vector, $p_\mu$ and the optical path length.

For CMB photons, even a very small coupling term can compound into a measurable rotation as they travel from $z \approx 1100$ to us today.  This would generate the forbidden $B$-mode correlations by rotating the primordial $E$-modes to $B$-modes as
\begin{eqnarray}
C_{\ell}^{TB} &=& C_{\ell}^{TE} \sin(2\alpha) \nonumber \\
C_{\ell}^{EB} &=& \frac{1}{2}C_{\ell}^{EE} \sin(4\alpha).
\label{eq:TBEB}
\end{eqnarray}

Current CMB measurements combined with astrophysical probes have constrained this overall rotation angle to be less than $1^\circ$ in magnitude \cite{Kaufman2014, Galaverni2014}.  However, this overall polarization rotation is completely degenerate with a polarization orientation miscalibration.  Using the suggested higher-precision measurements of the Crab Nebula the \PB\ telescope would be able to detect a $1^\circ$ polarization rotation at greater than 4$\sigma$ significance.

\section{Discrepancy Between CMB Polarization \& the Crab Nebula}

If we assume there is no cosmic polarization rotation, we can enforce that $C_{\ell}^{TB}$ and $C_{\ell}^{EB}$ must be zero and use Equation \ref{eq:TBEB} to ``de-rotate" CMB data, effectively using the CMB as a calibrator \cite{KSY2013}.  This self-calibration process has been used by several experiments, including \PB, \bicep1, \bicep2, \emph{Keck Array}, and ACTPol \cite{PB2014, B22014, Kaufman2013, Keck2015, ACTPol2014}.  However, using this method prevents measurements of cosmic polarization effects, and would yield incorrect results if CPR exists.

There is a discrepancy between telescope polarization orientation angles derived from observations of the Crab Nebula and self-calibration of approximately $1^\circ$ \cite{PB2014, ACTPol2014}.  More precise measurements of the Crab Nebula over multiple frequencies would allow us to resolve this discrepancy by measuring the variation of the polarization angle versus frequency with high precision.

\section{Example Observation}

Using the VLA in the C configuration, one can generate high-precision maps of the Crab Nebula in the K$_a$ and Q bands (33 and 44 GHz, respectively).  It would require 95 (169) pointings to cover the extent of the Crab Nebula in the K$_a$ (Q) band.  To achieve a sensitivity of 75 (40) $\mu$Jy/beam would require an on-source time of 18.7 seconds (3.6 minutes) per pointing to reach a statistical uncertainty of 0.01$^\circ$ on polarization angle.  This requires an observing time of $\sim$10 hours for the longest band.  As in Ref.~\refcite{VLAMars}, to achieve a 0.1$^\circ$ accurate measurement of the polarized emission of Mars would require only a few minutes of observations.

\section{Conclusions}


To achieve the $5\sigma$ significance detection level, the ``gold standard" in particle physics, for Inflationary gravitational waves at the $r=0.01$ level, we require a five-fold improvement in polarization calibration accuracy.  
We can take advantage of radio and microwave telescopes such as the VLA, ATCA, and ALMA, which can utilize the well-known polarized emission from Mars to measure the Crab Nebula's polarized emission to $0.1^\circ$ accuracy.
If the angles are constant in frequency, we can use these higher precision measurements to significantly improve current and near-future CMB telescopes' polarization orientation calibrations, necessary for investigations of Inflationary $B$-mode science.  If the Crab Nebula's polarization angles change with frequency, these measurements will be used to improve the calibration of existing \PB, QUIET, ACTPol, \emph{Planck} LFI data, and future Simons Array, CLASS, and Advanced-ACTPol data by a factor of five.  These measurements will be crucial for resolving the discrepancy between CMB-derived and Crab Nebula-derived polarization orientation angles, allowing us to constrain critical aspects of the standard model of cosmology, and beyond.


\section*{Acknowledgments}
The authors wish to thank Grant Teply for his insights, as well as the \bicep2 and \PB\ collaborations for useful discussions and feedback.

\bibliographystyle{ws-ijmpd}
\bibliography{VLA_references}

\end{document}